\documentclass{iau} 
\usepackage{graphicx}

\title[A critical assessment of models for multiple populations] 
{A critical assessment of models for the origin of multiple populations in globular clusters}

\author[Nate Bastian]   
{Nate Bastian$^1$}

\affiliation{$^1$Astrophysics Research Institute, Liverpool John Moores University, \\ 146 Brownlow Hill, Liverpool L3 5RF, UK \\ email: {\tt N.J.Bastian@ljmu.ac.uk} \\[\affilskip]}

\pubyear{2015}
\volume{316}  
\jname{Formation, evolution, and survival of massive star clusters}
\editors{A.C. Editor, B.D. Editor \& C.E. Editor, eds.}
\begin{document}

\maketitle

\begin{abstract}
A number of scenarios have been put forward to explain the origin of the chemical anomalies (and resulting complex colour-magnitude diagrams) observed in globular clusters (GCs), namely the AGB, Fast Rotating Massive Star, Very Massive Star, and Early Disc Accretion scenarios.  We compare the predictions of these scenarios with a range of observations (including young massive clusters (YMCs), chemical patterns, and GC population properties) and find that all models are inconsistent with observations.  In particular, YMCs do not show evidence for multiple epochs of star-formation and appear to be gas free by an age of $\sim3$~Myr.   Also, the chemical patterns displayed in GCs vary from one to the next in such a way that cannot be reproduced by standard nucleosynthetic yields.  Finally, we show that the ``mass budget problem" for the scenarios cannot be solved by invoking heavy cluster mass loss (i.e. that clusters were 10-100 times more massive at birth) as this solution makes basic predictions about the GC population that are inconsistent with observations.  We conclude that none of the proposed scenarios can explain the multiple population phenomenon, hence alternative theories are needed.
\end{abstract}

\firstsection 
\section{Introduction}

Multiple populations are now established as a near-ubiquitous property of globular clusters in the Galaxy (e.g., Gratton et al.~2012) as well as within nearby galaxies where their GCs can be resolved into individual stars (Mucciarelli et al.~2011; Larsen et al. 2012).  Abundance differences between stars in light elements such as He, Na, C, N, O and Al (within an individual cluster) can be found spectroscopically, and their influence on the lifetimes, bolometric corrections and colours of the stars results in complex colour-magnitude diagrams of the clusters.  This can be seen as split or spread main sequences, turn-offs, red giant branches and complex horizontal branch morphologies, depending on the filter combinations used (e.g, Sbordone et al. 2011).

Stars that are enriched in He, Na, N \& Al while being depleted in C and O are often referred to as ``2nd generation" (2G) stars, while the stars that have the abundance patterns observed in halo field stars of the same metallicity are often referred to as ``1st generation" (1G) stars.  This is somewhat misleading, as this is due to the interpretation of the patterns as due to multiple generations of star formation within the cluster, which is now looking as an unlikely explanation for the anomalies.  One point to note is that while age differences between the populations is possible (up to a few hundred Myr - e.g., Nardiello et al.~2015), age differences are not the reason for the complex CMDs.

A number of scenarios have been put forward in order to explain the origin of the multiple populations within GCs.  Most of the scenarios invoke multiple generations of star-formation within globular clusters, where a 2nd generation of stars is formed from material processed through first generation stars (i.e., the stellar ejecta of massive stars or AGB stars).  The models differ as to where the processed material comes from in order to form the 2G. The main candidates are Fast Rotating Massive Stars (FRMS), AGB stars, and Very Mass stars ($>10^4~M_{\odot}$).  The FRMS and AGB scenarios have been developed in some detail and examples of the scenarios can be found in Decressin et al.~(2007); D'Ercole et al.~(2008);  Conroy~(2012); and Krause et al.~(2013).

An additional constraint, not discussed in the present work, is that most GCs show small or absent spreads in Fe and other heavy elements.  Hence, any successful scenario will need to explain this lack of variation, in particular, how SNe ejecta did not pollute any of the gas if multiple generations of star-formation is invoked (c.f., Renzini~2013). 

These scenarios differ in many of their details but make some of the same predictions.  In the present work we will compare three of their basic predictions to observations.  1) The models do not invoke any special physics that only operates in the early Universe, so they should be happening in young massive clusters (YMCs) forming today.  2) Since they use the yields of evolved stars, there should be clear chemical patterns (modulo dilution with 'primordial' gas) observed in GC stars (i.e. relations between Na enhancement, O depletion and He spreads).  3) They all suffer from a ``mass budget problem" where there is simply not enough material being processed through 1G stars to form the observed number of 2G stars.  In order to solve this, these models all suggest that GCs were much more massive at birth (by factors of 10-100 times their present day masses), and invoke heavy mass loss (through gas expulsion or galactic tides) to explain the current fractions of enriched stars.  This heavy mass loss should leave a clear imprint on the GC cluster population.

Alternatively, Bastian et al.~(2013a) suggested a scenario that only entails a single generation of stars to form, and that low-mass stars with protoplanetary discs sweep up material shed from interacting binary or rapidly rotating massive stars. This is referred to as the Early Disc Accretion (EDA) scenario and like the above scenarios, it predicts clear chemical patterns for GC stars today.

In the following sections we compare the models to a variety of observations.  No model is found that can satisfy the observational constraints.

\section{Constraints from Young Massive Clusters}

Of course GC formation at high redshift may have been entirely different than YMC formation happening locally.  However, none of the models put forward so far invoke any special physics applicable only to the early Universe, so observations of YMCs can be used to directly test the proposed models.  There have been a number of recent works on this subject which we summarise here:

\begin{itemize}

\item {\bf A lack of ongoing star-formation within YMCs:} Bastian et al. (2013b) studied $\sim130$ Galactic and extragalactic YMCs with ages between 10-1000~Myr and masses between $10^4 - 10^8~M_{\odot}$ looking for evidence of ongoing star-formation within them based on emssion lines. No evidence for secondary star-formation events where found, which is in conflict with predictions of the AGB scenario.
\item {\bf The star-formation history (SFH) of a YMC:} Cabrera-Ziri et al.~(2014) used a high S/N optical spectrum of NGC~34-cluster 1 to estimate its SFH.  The cluster was found to be best fit by a single age populations (age$\sim100$~Myr; mass$\sim10^7~M_{\odot}$).  This is in conflict with predictions of the AGB scenario.
\item {\bf CMDs of YMCs:} Larsen et al.~(2011) and Niederhofer et al.~(2015a) studied resolved CMDs of YMCs (with masses between $10^4-10^6~M_{\odot}$ and ages between a few Myr and 1~Gyr) in comparison with stellar isochrones (essentially working out the SFH of the clusters).  While age spreads could not be ruled out for all clusters (see also Niederhofer et al.~2015b) none were consistent with the age spreads expected in the AGB scenario.
\item {\bf Young gas free clusters:} Hollyhead et al.~(2015) and Bastian et al.~(2014) studied young ($<10$~Myr) massive clusters and found that they were gas free with $\sim3$~Myr of their formation.  This is in conflict with the FRMS scenario which requires left-over gas from the formation of the 1G stars to remain in the cluster for $7-20$~Myr.
\item {\bf A lack of gas in older clusters} Longmore~(2015) showed that in the AGB scenario (as detailed in D'Ercole et al.~2008) clusters with ages between 30-100~Myr should have large amounts of gas within them, which will result in heavy extinction within the cluster, even out to 3-5~pc from the cluster centre.  This is in basic disagreement with observations of YMCs which often find low extinction for clusters in this age range.
\item {\bf ALMA observations of YMCs:} Similarly, Cabrera-Ziri et al.~(2015) used ALMA to search for gas in and around YMCs in the Antennae galaxies (ages$\sim50-200$~Myr, mass$\sim10^6~M_{\odot}$) which is predicted to remain in or near the clusters.  No such gas was found, down to limits of $<10\%$ of the stellar mass, in conflict with the AGB model.

\end{itemize}

{\bf Observations of massive stellar clusters forming today are inconsistent with the AGB and FRMS scenarios.  At the very least the proposed scenarios would not be generally applicable to all massive clusters, but instead would need some additional criteria that is only in operation in the early Universe.  At worst, the basic idea of multiple generations of star-formation within massive clusters is not valid.}

\section{Constraints from Chemical Abundance Patterns}

All of the proposed models use the yields of certain stars (AGB, FRMS, VMS, or interacting binary stars), together with the abundance patterns observed in the 'primordial' population (i.e. those stars with abundance patterns similar to the field) to make predictions for the expected trends in certain elements.  For example, by diluting the yields of AGB stars with gas with the abundance patterns of the primordial stars, one expects an anti-correlation between Na and O (whereas the yields themselves predict a correlation).  This also applies to other elements, so we can test, for example, if the predicted spreads in Na and O are consistent with the observed spreads in He.  This is a key constraint, as basic nucleosynthesis and stellar evolution predict clear correlations between these three elements.  See Bastian et al.~(2015) for a further description of the tests and expectations.

\begin{figure}[b]
\begin{center}
 \includegraphics[width=5.5in]{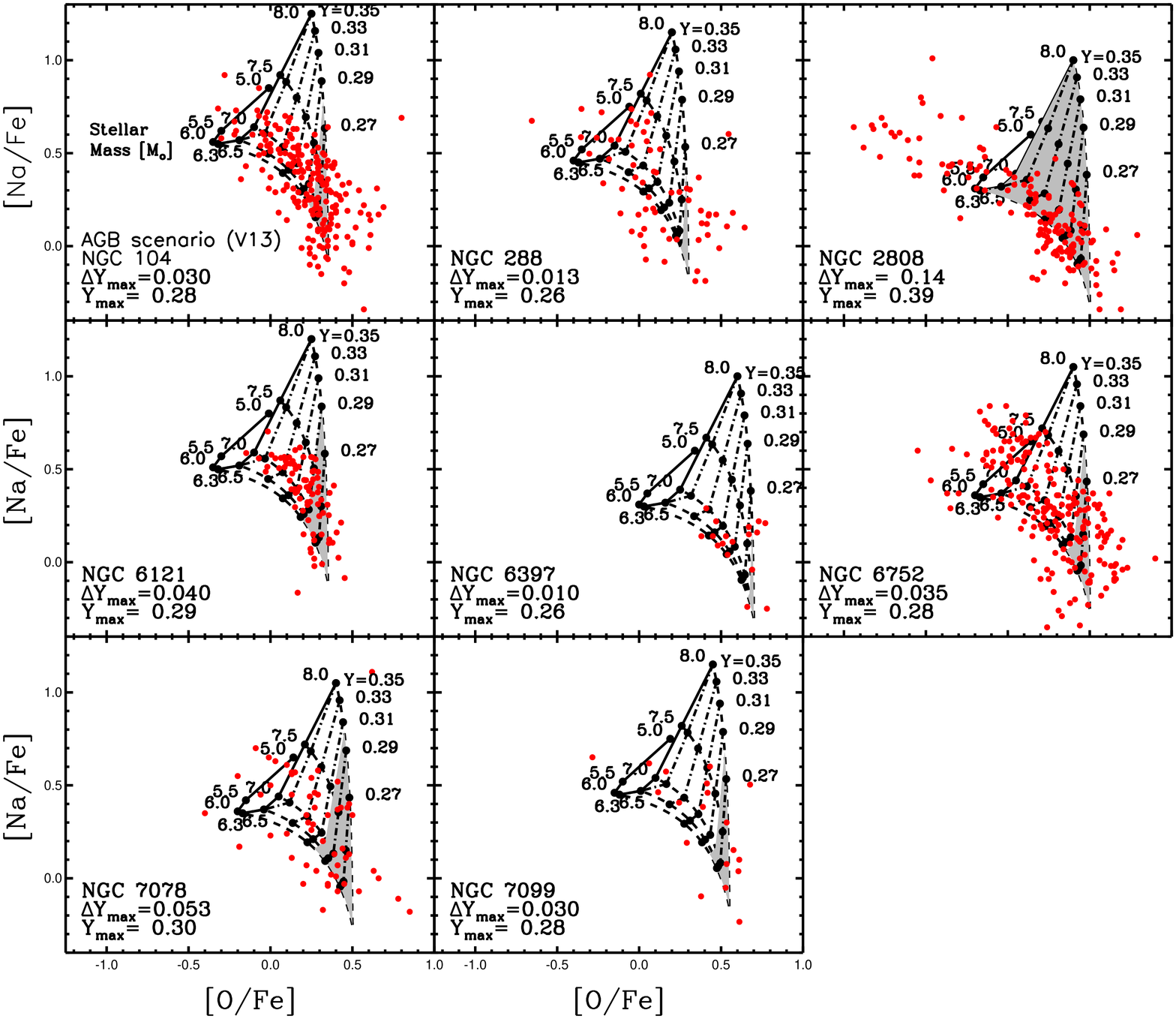} 
 \caption{Dilution models for the AGB scenario using the yields of Ventura et al.~(2013).  Solid lines show the yields, while dashed lines show the dilution model adopting the yields for each stellar mass.  Lines of constant He enrichment are shown as dash-dotted lines.  For a given He spread ($\Delta$Y) the allowed range of Na-O values for these yields is shown as a shaded region.  Note that all clusters violate this constraint.  The same applies to all AGB yields in the literature as well as the yields of rotating massive stars, interacting massive binaries or very massive stars.  Taken from Bastian et al.~(2015).}
   \label{fig1}
\end{center}
\end{figure}

In Fig.~\ref{fig1} we show such dilution models for the AGB scenario, using the yields of AGB stars from Ventura et al.~(2013).  These represent some of the most optimistic yields for AGB stars if they are to be the origin of the anomalies seen in GCs \footnote{Compare these yields with the AGB yields of Doherty et al.~2014, who conclude that AGB stars cannot be the origin of the chemical anomalies in GCs.}.  The solid lines show the yields for the AGB stars of different masses, and the dashed lines show the dilution model for each mass.  The dash-dotted lines show regions of constant He.  So, for a given position in the Na-O plane, one should be able to explicitly predict the He abundance for that star (relative to the ``primordial" stars).  By looking at the most extreme stars in the Na-O plane, the maximum He spread for that cluster can be estimated. The filled (red) circles show data for stars within a given GC. Finally, for each GC, the allowed range in the Na-O plane for cluster stars, for the measured He spread within the cluster, is shown as a filled region.

It is clear that all the GCs have significant numbers of stars outside the allowed range.  In fact, most of the GCs have similar spreads in Na and O, while having extremely different He contents.  Hence, these specific yields are not consistent with the observations.  We have also compared the observations (not shown) with the yields of FRMSs (Decressin et al.~2007), a number of other available AGB models, Very Massive Stars (Denissenkov 
\& Hartwick~2014), and massive interacting binaries (de Mink et al.~2009).  All of these sources are at odds with observations, with the general problem that they all over-predict the amount of He spread for a given Na-O spread.  {\bf Hence, none of the yields put forward in the literature so far are consistent with the observations.}

Additionally, the yields of AGBs and FRMS are expected to be metallicity dependent, so that trends with the type of Na-O spreads observed should vary as a function of metallicity.  To date, no such variations have been found (e.g., Bastian et al.~2015).

However, the problem is in fact much worse than the yields being wrong.  Bastian et al.~(2015) also made empirical yields, by taking the observed Na-O spread of a given cluster and associating this with the maximum allowed spread in He.  By comparing the ``empirical yield" from each cluster to others, it is clear that all clusters are different, i.e. if a hypothetical source could explain the trends in one cluster, say 47 Tuc, they would not be able to fit any other cluster, for example, NGC~2808.  

Hence, a large range in stochasticity is required in the yields, if nucleosynthesis is the source of the enriched material.  However, none of the suggested sources (AGBs, FRMS, etc) will be able to produce the required amount stochasticity. 

{\bf The need for large amounts of stochasticity in the source of the enriched material  effectively rules out all the proposed sources (e.g., AGB, FRMS, interacting massive binaries, VMS, etc).}

\section{Constraints from Globular Cluster Population Properties}

Next we consider constraints on the models from the properties of the GC population in general.  As discussed in the introduction, scenarios that invoke the formation of a second generation of stars from the ashes (i.e. nuclear processed material) of a first generation all suffer from a severe mass budget problem.  This is because AGB or massive stars contribute very little to the overall mass of the cluster (due to the stellar IMF), so they can only process a rather limited amount of material.  On the contrary, enriched stars are observed to make up $>50$\% of stars within GCs (whereas the expected fraction would be $1-10$\% depending on the source and efficiency of the process).  In order to solve this basic mass budget problem, it has been suggested that GCs were much more massive at birth and have preferentially lost most of the 1G stars within them.  

In order to do this, the scenarios have invoked cluster expansion due to stellar mass loss in a mass segregated cluster.  As the clusters expand, their outer regions will be lost due to galactic tides (as the 2G stars are expected to be more centrally concentrated), changing the enriched fraction (F$_{\rm enrich} = N_{\rm enriched} / N_{\rm total}$) from the initial value of $\sim0.05$ to the observed value of $>0.5$.  This can only work if the galactic tidal field is strong enough, and due to this, scenarios that invoked this mechanism have put the young GCs at small Galactocentric radii ($\sim4$~kpc) in order to maximise the loss of 1G stars.

\begin{figure}[b]
 \includegraphics[width=5.5in]{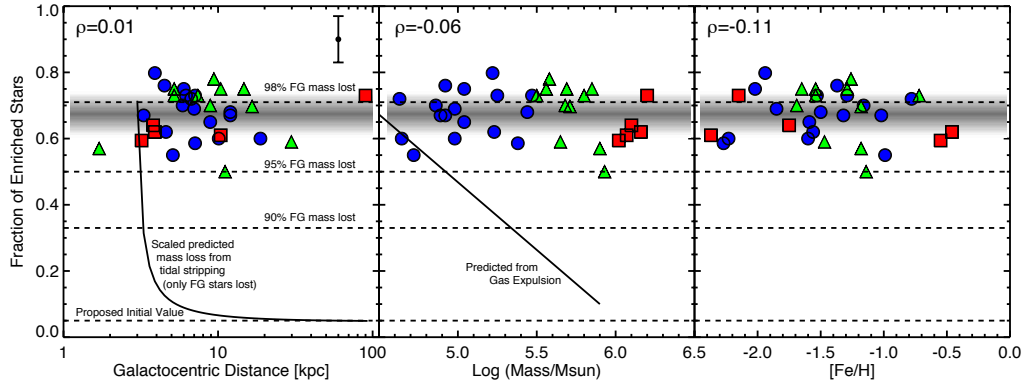} 
 \caption{The fraction of enriched stars (F$_{\rm enrich}$) as a function of Galactocentric distance, GC mass and metallicity in a sample of 34 Galactic GCs.  The lines in the left and centre panels denote predictions of mass loss mechanisms that have been invoked to make GCs lose significant amounts of stellar mass in order to solve the ``mass budget problem".  The constancy of F$_{\rm enrich}$ is inconsistent with any of the scenarios suggested so far.  Taken from Bastian \& Lardo~(2015).}
   \label{fig_enriched}
\end{figure}

A clear prediction of this kind of GC dissolution is that clusters that experience weaker tidal fields (e.g., clusters at larger Galactocentric radii) should have a lower fraction of F$_{\rm enrich}$, tending towards the initial value.  Bastian \& Lardo~(2015) have searched for such a relation using 34 GCs with abundance measurements (or HST imaging that can separate the different populations within GCs with certain filter combinations) and the results are shown in Fig.~\ref{fig_enriched}.  The data show no correlation between F$_{\rm enrich}$ and Galactocentric distance, while the expected relation in such a heavy mass loss scenario is shown as a solid line. In fact, the data are consistent with a constant F$_{\rm enrich} = 68\pm7$\%.  The predictions clearly inconsistent with this, suggesting that this mass-loss mechanism cannot be responsible for the heavy mass loss required in such multiple-generation scenarios.

Khalaj \& Baumgardt~(2015) have studied another potential mass loss mechanism, the rapid expulsion of gas from young GCs, which also causes the clusters to expand.  This type of mass loss is also expected to result in a strong relation between F$_{\rm enrich}$ and Galactocentric distance, but also a strong relation with GC mass.  This is tested in the middle panel of Fig.~\ref{fig_enriched}, and is again incompatible with the data, in agreement with the conclusions of Khalaj \& Baumgardt~(2015).  Finally, the in right panel we show the observed relation between F$_{\rm enrich}$ and GC metallicity, and again there is no correlation.  This suggests that the location and age of the GC does not affect the fraction of enriched stars.

Finally, Kruijssen~(2015) have shown that because the tidal environment experienced by a GC is dependent on the galaxy host of which it forms, which in turn is related to the metallicity of the host galaxy (due to the mass-metallicity relation), F$_{\rm enrich}$ would be expected to be low for low-metallicity GCs and increase for high-metallicity GCs.  This is clearly not compatible with the data, as shown in the right panel of Fig.~\ref{fig_enriched}.  The author concludes that GCs were, on average, only a factor of $2-3\times$ more massive at birth than they are presently.

{\bf Taken together, this suggests that if GCs did undergo strong mass loss, they must all have lost between 95 and 99\% of their initial masses (assuming only 1G stars are lost) independent of the mass, Galactocentric distance, mass, age, location at birth or subsequent migration.  No mass loss mechanism proposed to date can match these criteria.  Hence, we conclude that GCs did not in fact undergo heavy mass loss and that the observed fraction of enriched stars is representative of their initial value.  This effectively rules out standard nucleosynthesis as the origin of the chemical anomalies.}

\section{Other constraints on the Mass Budget}

Larsen et al.~(2012; 2014) have investigated the GC properties in a sample of three dwarf galaxies, namely the Fornax, WLM and IKN dwarf galaxies.  These galaxies have very high specific frequencies (the fraction of light in GCs relative to the field), especially at low metallicities.  The authors found that below [Fe/H]$=-2$, GCs make up between 10-30\% of the total stellar mass of the galaxy.  This is incompatible with the idea of heavy cluster mass loss, as the stars lost from the GCs would share the same metallicity as the GCs and make up a substantial fraction of the field.  Fornax, in particular, is a very isolated galaxy and is not expected to have undergone any strong interactions which may have resulted in the loss of stars from the galaxy.  In any case, it is very improbably to strip field stars from a galaxy without stripping GCs as well.  

Schiavon et al.~(2015) have found, based on the APOGEE survey of stars in the Galactic bulge, a relatively high percentage of ``enriched" stars in the field, relative to primordial stars.  Under the extreme assumption that the all the field stars are from dissolved GCs, the authors find a lower limit to F$_{\rm enrich}$ of 0.07, in tension with model predictions.  Under more realistic assumptions, this value increases substantially.

The observations by Larsen et al. and Schiavon et al.  directly contradict a basic (and necessary) premise of scenarios that invoke multiple generations of star-formation within GCs, that heavy loss of 1G stars is required to solve the ``mass budget problem".

Finally, it has been found that integrated [N/Fe] of GCs correlates with GC mass in M31 (Schiavon et al.~2013; Colucci et al.~2014; and likely in the Galaxy as well) and that the spread of He ($\Delta$Y) within Galactic GCs also is proportional to mass (Milone~2015).  This is inconsistent with self-enrichment scenarios, as the amount of N or He produced should be constant per unit stellar mass (as the IMFs are fully sampled).  In order to accommodate the observations, low mass clusters must have lost substantial fractions of the total amount of material processed through 1G stars.  This further exacerbates the mass budget problem by a factor of 3 or more (Bastian et al.~2015).

\section{Summary}

The constraints presented here effectively rule out all the main scenarios for the formation of multiple populations within GCs.  First, we have shown that observations of young massive clusters are incompatible with predictions from the FRMS and AGB scenarios, so at the very least an additional component is needed in these models to make them only operate in the early Universe, and at worst this shows that the basic idea behind multiple generations of stars forming in the same cluster (outside nuclear clusters) is not valid.  

Secondly, none of the proposed nucleosynthetic sources can produce the observed chemical patterns in GC stars.  They cannot 1) produce the observed variations from GC to GC (i.e., a large amount of stochasticity is required that none of the sources can provide) and 2) all sources predict a strong correlation between the extent of the Na-O anti-correlation and the He spread within the GC, in contrast with observations.  

Finally, the observed fraction of enriched stars within GCs is remarkably constant, $F_{\rm enrich} = 68\pm7$\%, independent of a cluster's mass, galactocentric distance, metallicity, location of birth or subsequent migration.  This is incompatible with basic predictions of a scenario where GCs preferentially lost a large fraction of their initial mass (i.e., lost large amounts of FG stars).  Hence, the mass budget problem cannot be solved by invoking a scenario where GCs lost large amounts of mass, confirming previous results from studies of GCs and field stars in dwarf galaxies.

The chemical patterns observed in GCs, when combined with limits on the mass budget of GCs, rule out standard nucleosynthesis the process responsible for the observed abundance anomalies.  Hence, with the exclusion of all current models, new scenarios are desperately needed.  

\begin{figure}[b]
 \includegraphics[width=3.65in,angle=270]{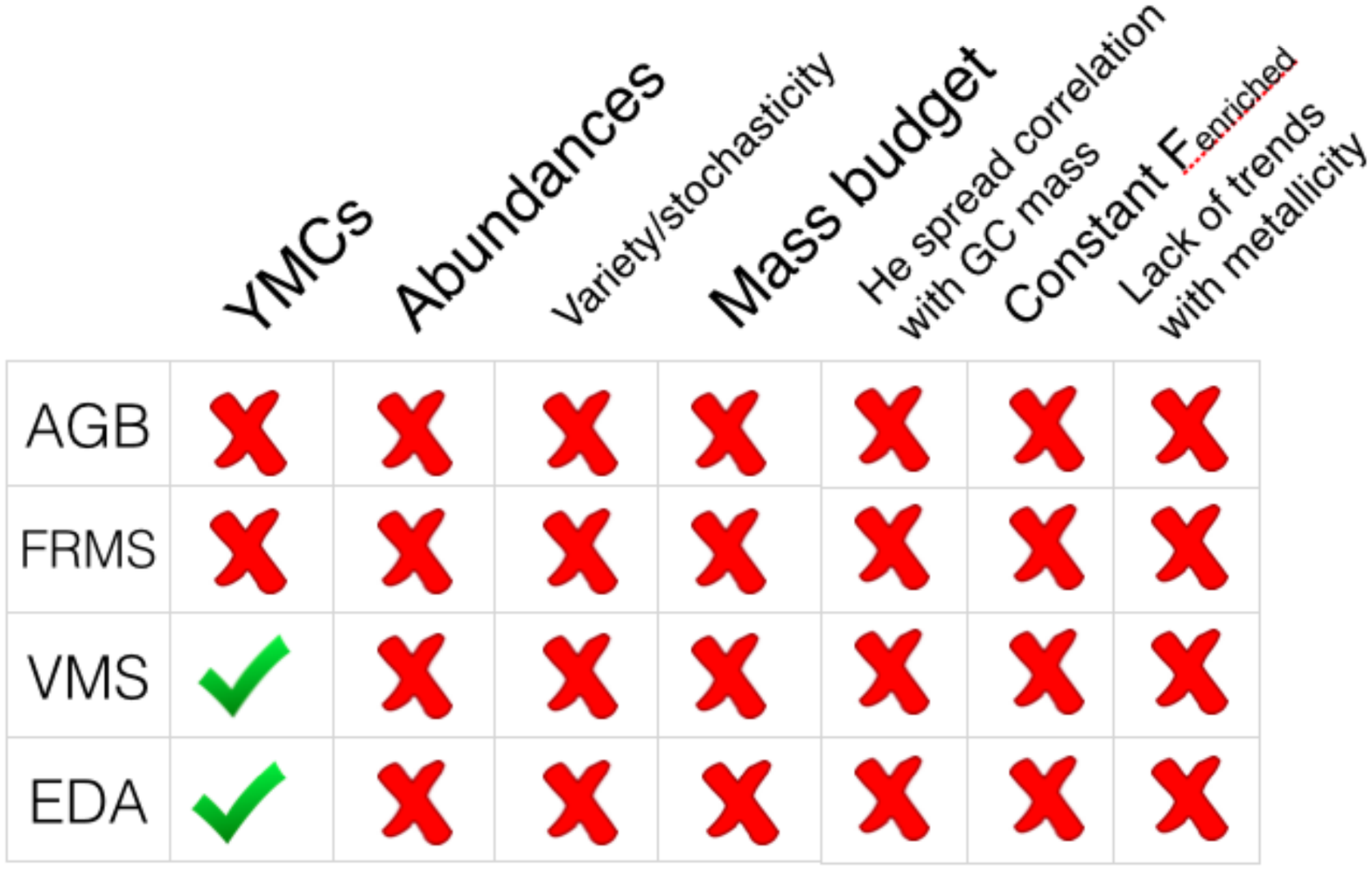} 
 \caption{A summary of the results presented in the current work.  A green check means that a given constraint does not directly violate a prediction/requirement of a given scenario, while a red cross means that the observational constraints directly contradict a prediction/requirement of a scenario.}
   \label{fig_table}
\end{figure}

We conclude with a visual summary of the constraints discussed in the present work shown in Fig.~\ref{fig_table}.

\end{document}